\documentclass[sigplan,screen]{acmart}

\usepackage{subcaption}
\usepackage{booktabs}
\usepackage{dsfont}
\usepackage{xspace}
\usepackage{xcolor}
\usepackage{algorithm}
\usepackage{algpseudocodex}
\usepackage{multirow}
\usepackage{url}

\setcopyright{none}
\renewcommand\footnotetextcopyrightpermission[1]{}

\newcommand{\reffig}[1]{Figure~\ref{#1}}
\newcommand{\reftab}[1]{Table~\ref{#1}}
\newcommand{\refsec}[1]{\S\ref{#1}}

\newcommand{\refequ}[1]{(\ref{#1})}

\newcommand{\refalg}[1]{Algorithm~\ref{#1}}

\newcommand{\mech}{Hydra\xspace}
\newcommand{\dt}{DT}
\newcommand{\rf}{RF}
\newcommand{\svd}{SVD}
\newcommand{\reptree}{RT}
\newcommand{\mlp}{MLP}
\newcommand{\ens}{Ens\xspace}

\definecolor{fs1}{HTML}{1f77b4}
\definecolor{fs2}{HTML}{aec7e8}
\definecolor{fs3}{HTML}{ff7f0e}
\definecolor{fs4}{HTML}{ffbb78}
\definecolor{fs5}{HTML}{2ca02c}
\definecolor{fs6}{HTML}{98df8a}
\definecolor{fs7}{HTML}{d62728}
\definecolor{fs8}{HTML}{ff9896}
\definecolor{fs9}{HTML}{9467bd}
\definecolor{fs10}{HTML}{c5b0d5}
\definecolor{fs11}{HTML}{8c564b}

\begin{document}

\title{\mech: Robust Hardware-Assisted Malware Detection}

\author{Eli Propp}
\email{eli.propp@uwaterloo.ca}
\affiliation{
       \institution{University of Waterloo}
       \city{Waterloo}
       \country{Canada}
}

\author{Seyed Majid Zahedi}
\email{smzahedi@uwaterloo.ca}
\affiliation{
       \institution{University of Waterloo}
       \city{Waterloo}
       \country{Canada}
}

\begin{abstract}
Malware detection using Hardware Performance Counters (HPCs) offers a promising, low-overhead approach for monitoring program behavior. However, a fundamental architectural constraint, that only a limited number of hardware events can be monitored concurrently, creates a significant bottleneck, leading to detection blind spots. Prior work has primarily focused on optimizing machine learning models for a single, statically chosen event set, or on ensembling models over the same feature set. We argue that robustness requires diversifying not only the models, but also the underlying feature sets (i.e., the monitored hardware events) in order to capture a broader spectrum of program behavior. This observation motivates the following research question: \emph{Can detection performance be improved by trading temporal granularity for broader coverage, via the strategic scheduling of different feature sets over time?} To answer this question, we propose \mech, a novel detection mechanism that partitions execution traces into time slices and learns an effective schedule of feature sets and corresponding classifiers for deployment. By cycling through complementary feature sets, \mech mitigates the limitations of a fixed monitoring perspective. Our experimental evaluation shows that \mech significantly outperforms state-of-the-art single-feature-set baselines, achieving a $19.32\%$ improvement in F1 score and a $60.23\%$ reduction in false positive rate. These results underscore the importance of feature-set diversity and establish strategic multi-feature-set scheduling as an effective principle for robust, hardware-assisted malware detection.
\end{abstract}

\pagestyle{fancy}
\fancyhead{}
\fancyfoot{}

\maketitle
\thispagestyle{empty}

\section{Introduction} \label{sec:introduction}

Malware continues to evolve in scale and sophistication, posing persistent threats to modern computing infrastructures and critical services. As malicious software increasingly targets heterogeneous and large-scale systems, the design of accurate and efficient malware detection mechanisms remains a foundational challenge for both security research and operational deployments.

Traditional signature-based detection remains widely deployed due to its efficiency and low false-positive rates on known threats. To extend coverage beyond signatures, prior work has explored both static and dynamic analysis techniques. Static analysis extracts structural features such as control-flow graphs and opcode sequences without executing the program, whereas dynamic analysis monitors runtime behavior in controlled environments by observing API calls, network activity, and system state changes.

As a complementary approach, prior work has investigated dynamic malware detection using low-level hardware events such as cache misses and branch mispredictions~\cite{malone2011hardware, zhou2018hardware, demme2013feasibility, basu2019theoretical, cheng2023feasibility}. This line of work exploits program \emph{phase behavior}, where execution transitions through distinct stages with characteristic microarchitectural signatures, enabling discrimination between benign and malicious workloads. While similar Hardware Performance Counter (HPC) profiling is widely used for performance analysis, malware detection faces the additional challenge that only a small number of events (often $4$--$8$) can be monitored concurrently.

Most hardware-assisted malware detection systems follow a common pipeline: executables are run in controlled environments while HPCs collect traces from multiple events, informative features are selected from these traces, and a machine learning (ML) model is trained on the resulting feature set. Although prior studies report promising results, subsequent analyses have raised concerns about robustness, showing substantial performance variability and high false discovery rates~\cite{zhou2018hardware}.

Building on these observations, we argue that detection performance depends \emph{jointly} on the choice of learning model and the specific hardware events being monitored. Relying on a static event set fails to account for the fact that different programs may leave defining signatures in different subsets of events. As a result, a feature set optimized for the average case may introduce \emph{blind spots}, where unwanted activity goes undetected or benign activity is incorrectly flagged.

A natural response is to use ensemble methods~\cite{sayadi2018ensemble,sayadi20192smart}, where multiple detection models of the same or different architectures are combined. While ensembling models trained on the same events can improve performance given identical evidence, it cannot recover evidence that was never measured. We therefore argue that diversifying \emph{what is measured} by utilizing multiple feature sets is necessary to reduce these blind spots and achieve more robust detection.

We illustrate this intuition with an analogy. Consider the task of detecting motion in a room when only a few sensors can be active at once. Sensor placement corresponds to the event set, and the limited number of active sensors mirrors the HPC constraint. Placing multiple sensors at the same location is analogous to ensembling multiple detection models over the same feature set: although redundancy may reduce noise, the shared viewpoint may still fail to detect motion near the edges of the room. By contrast, distributing sensors across the room provides broader coverage. Similarly, using multiple feature sets offers wider behavioral visibility and improves robustness.

However, the fundamental restriction remains: only a single set of features can be monitored simultaneously. Given this counter budget, we ask the following research question: \emph{can detection accuracy and robustness be improved by trading temporal granularity for broader coverage, by strategically scheduling different feature sets sequentially over time?}

To address this question, we propose \mech, an end-to-end mechanism for utilizing multiple feature sets in feature-restricted malware detection using HPCs. We address the feature restriction by splitting each execution trace into time-ordered slices, allowing \mech to employ different feature sets across slices. \mech then learns, offline, an effective schedule of feature sets and ML models for deployment.

This approach introduces an inherent trade-off. Because decisions are made at the slice level, each model has access to less information, which can reduce per-slice predictive performance. Nevertheless, by strategicaly scheduling predictions across complementary feature sets and leveraging ensembling, \mech can offset this information loss with broader and more diverse evidence.

Through experiments, we show that \mech learns an effective schedule of feature sets that outperform state-of-the-art, single-feature-set baselines, achieving a $19.32\%$ improvement in F1 score and a $60.23\%$ reduction in false positive rate relative to the top-performing single-feature-set baseline.

The remainder of the paper is organized as follows. In \refsec{sec:background}, we summarize common mechanisms for traditional and HPC-based malware detection. \refsec{sec:methodology} outlines our data processing pipeline and baseline implementations. We then formally introduce \mech in \refsec{sec:mechanism}. \refsec{sec:eval} presents our experimental setup, evaluation of \mech, and sensitivity analyses. Finally, we conclude in \refsec{sec:conclusion}.

\section{Background and Related Work} \label{sec:background}

This section, we begin by summarizing traditional and contemporary detection methods before focusing on the use of low-level hardware events. We detail the standard pipeline employed by prior hardware-based detectors, including data collection, feature selection, and model construction. Finally, we examine the limitations of current approaches, thereby motivating the need for a more robust detection strategy, which we detail in subsequent sections.

\subsection{Malware Detection}\label{subsec:traditional}

Traditional signature-based detection remains one of the most widely deployed approaches in both commercial and academic settings, owing to its efficiency and low false-positive rates for known threats. This method maintains a database of byte-level or hash-based signatures extracted from previously identified malware samples and compares them against incoming binaries at scan time, forming the foundation of widely deployed detection engines such as \texttt{ClamAV}~\cite{clamav}.

Beyond signature-based approaches, prior work has investigated a range of complementary techniques, including static and dynamic analysis. Static binary analysis examines program structure without execution, leveraging features such as control-flow graphs and opcode sequences to train machine learning classifiers~\cite{anderson2018ember}. In contrast, dynamic execution analysis observes runtime behavior within controlled environments (e.g., sandboxes or virtual machines), monitoring API calls, network activity, and system state changes to identify anomalies indicative of malicious activity~\cite{or2019dynamic}.

\subsection{Hardware Events for Behavioral Profiling}

Prior work has explored the use of low-level hardware events for dynamic execution analysis~\cite{malone2011hardware, sayadi2018ensemble, he2021machine, zhou2018hardware, demme2013feasibility, basu2019theoretical, bahador2014hpcmalhunter}. These events, such as branch mispredictions, cache misses, and Translation Lookaside Buffer (TLB) misses, are byproducts of program execution and are monitored in real time by modern processors. The fundamental premise is that program execution exhibits \textit{phase behavior}: a program transitions through distinct phases (e.g., initialization, computation, and I/O), each characterized by specific patterns of interaction with the microarchitecture. These phases consequently generate hardware events with distinctive frequencies and temporal structures. Because execution phases and their behavioral signatures are often characteristic of a given program (or class of programs), monitoring hardware event patterns could enable identification of malicious behaviors, which may differ substantially from benign execution.

This concept of hardware event profiling for behavioral characterization is not exclusive to security. Major technology companies like Netflix and Google routinely profile their workloads using hardware performance counters (HPCs) to understand performance characteristics, identify bottlenecks, and optimize resource utilization for efficiency and cost analysis~\cite{netflix-perf1,netflix-perf2,google-perf-vm1,google-perf-vm2}.

Low-level hardware events are tracked by special-purpose registers known as Hardware Performance Counters (HPCs)~\cite{mucci1999papi,uhsadel2008exploiting,weaver2008can}. While processors can typically monitor dozens of distinct event types, architectural constraints limit the number that can be monitored \textit{concurrently} to a small set, often between 4 and 8. This limitation necessitates a critical step in building a hardware-based detector: selecting the most informative subset of events to monitor.

\subsection{Prior Work: A Standard Pipeline}

The majority of prior work on hardware-based malware detection follows a common pipeline. First, a dataset containing both malicious and benign executable binaries is constructed. These binaries are then executed (often multiple times) in a controlled environment, while hardware performance counters (HPCs) are configured to collect traces from a broad set of hardware events.

Next, due to the limited number of events that can be monitored concurrently, feature selection and transformation algorithms are applied to the collected traces to identify the most informative hardware events for discriminating between malware and benignware. Common approaches include Principal Component Analysis (PCA)~\cite{kadiyala2020hardware, sayadi2018customized, zhou2018hardware, sayadi20192smart},
correlation-based attribute evaluation~\cite{sayadi20192smart}, recursive feature elimination~\cite{he2021machine}, and manual selection~\cite{ganfure2022deepware, malone2011hardware}.

Finally, a detection model is trained using the selected events as features. Prior studies have explored a wide range of machine learning (ML) models, including classical algorithms such as decision trees~\cite{kadiyala2020hardware, sayadi2018customized, demme2013feasibility, he2021machine}, random forests~\cite{kadiyala2020hardware, he2021machine}, multilayer perceptrons~\cite{sayadi2018customized, sayadi20192smart}, k-nearest neighbors~\cite{kadiyala2020hardware, demme2013feasibility}, and Gaussian Naive Bayes~\cite{he2021machine}. More complex neural network architectures have also been investigated~\cite{kadiyala2020hardware, ganfure2022deepware, alam2019ratafia}. Ensemble techniques such as AdaBoost~\cite{kadiyala2020hardware, he2021machine, sayadi2018ensemble} and bootstrap aggregating (bagging)~\cite{sayadi2018ensemble} are frequently employed to further improve detection performance.

\subsection{Limitations}

Despite demonstrations of effectiveness, the robustness of hardware-based malware detection has been challenged. Zhou et al.~\cite{zhou2018hardware} question the core assumption that malicious behavior consistently manifests in distinct microarchitectural activity. Their empirical study, involving extensive cross-validation of various ML models, revealed substantial performance variability and a high false discovery rate, casting doubt on the reliability of a single model trained on a fixed event set.

Building on these insights, we posit and empirically demonstrate that detection efficacy depends critically on two factors: the choice of the ML model \textit{and} the specific set of hardware events (features) it monitors. A detector trained on a fixed feature set inevitably develops \textit{blind spots}: malware samples whose execution patterns are insufficiently distinguishable from benign behavior (false negatives), as well as benign programs that are mistakenly flagged as malicious (false positives). These blind spots arise from inherent model limitations or, more fundamentally, from the insufficient discriminative power of the chosen events for certain malware families. Consequently, no single model or fixed feature set can guarantee comprehensive detection coverage.

Thus, statically deploying any single detector, even a high-performing one, leaves exploitable gaps. To enhance robustness, our approach, detailed in \refsec{sec:methodology}, strategically schedules multiple detectors, each trained on a different, complementary set of hardware events. This strategy aims to achieve superior overall detection performance by mitigating the individual blind spots of any single detector, moving beyond the limitations of a static, single-feature-set deployment.
\section{Baseline Detection Models} \label{sec:methodology}

This section details the methodology for constructing the baseline detection models that serve as foundational components for \mech (see \refsec{sec:mechanism}). An effective detection mechanism relies on robust, well-evaluated base classifiers. Our methodology encompasses three critical stages: the principled processing of microarchitectural trace data (\refsec{subsec:data-proc}), the strategic selection of hardware performance counter (HPC) events as discriminative features (\refsec{subsec:feat-sel}), and the evaluation of diverse machine learning classifiers to establish performance baselines (\refsec{subsec:baseline-model-perf}).

\begin{table}[t!]
    \centering
    \footnotesize
    \begin{tabular}{llll}
        \toprule
        \textbf{\#} & \textbf{Event} & \textbf{\#} & \textbf{Event} \\
        \midrule
        1 & \texttt{all-dc-accesses} & 12 & \texttt{cache-misses} \\
        2 & \texttt{l1-dc-ld} & 13 & \texttt{l1-dtlb-misses} \\
        3 & \texttt{ls-dc-accesses} &  14 & \texttt{ls-l1-dtlb-misses} \\
        4 & \texttt{ld-dispatch} & 15 & \texttt{l1-ic-ld-misses} \\
        5 & \texttt{branches} & 16 & \texttt{ls-mab-alloc-st} \\
        6 & \texttt{st-dispatch} & 17 & \texttt{dtlb-ld-misses} \\
        7 & \texttt{cache-accesses} & 18 & \texttt{l2-dtlb-misses} \\
        8 & \texttt{l1-dc-ld-misses} & 19 & \texttt{itlb-ld} \\
        9 & \texttt{l2-dc-accesses} & 20 & \texttt{l2-itlb-misses} \\
        10 & \texttt{l2-ic-accesses} & 21 & \texttt{itlb-ld-misses} \\
        11 & \texttt{branch-misses} &&\\
        \bottomrule
    \end{tabular}
    \caption{\textbf{Microarchitectural events}.}
    \label{tab:features}
\end{table}

\subsection{Data Processing} \label{subsec:data-proc}

Robust data processing is paramount for any machine learning (ML) pipeline, particularly in security applications where evaluation validity is critical. Our processing begins with a curated collection of benign and malicious executable samples, as detailed in \refsec{subsubsec:data-coll}. Each sample is executed in a controlled, instrumented environment to generate fine-grained execution traces that record the occurrence of predefined hardware events.

To render this temporal data suitable for conventional ML classifiers, we follow the established methodology of~\cite{demme2013feasibility, zhou2018hardware}. Execution traces are segmented into discrete, non-overlapping windows. For each window, we aggregate the frequency of each monitored hardware event, effectively transforming a multivariate time series into a single feature vector. Each aggregated window constitutes one training or testing sample. During inference, we classify each window independently. The final sample-level prediction is determined by a majority vote: if the majority of a sample's windows are predicted as malicious, the entire sample is classified as malware.

Constructing valid training and testing datasets for security tasks requires meticulous attention to avoid common pitfalls~\cite{arp2022and,pendlebury2019tesseract}. We address these as follows:
\begin{itemize}
    \item \textbf{Sampling Bias / Spatial Inconsistency}: To ensure our evaluation reflects a realistic operational scenario where benign software predominates, we construct our test set to contain 90\% benign samples, adhering to recommendations from prior work~\cite{pendlebury2019tesseract}.
    \item \textbf{Label Inaccuracy}: We verify all malware samples using \texttt{ClamAV}~\cite{clamav} to preclude benign samples from being mislabeled as malicious. While a negative \texttt{ClamAV} scan does not guarantee benignity, a positive classification provides a high-confidence malicious label.
    \item \textbf{Test Snooping}: We strictly enforce a separation between training and testing data, ensuring no information from any test sample (e.g., for feature selection or hyperparameter tuning) influences the model training process, thereby avoiding the P3 pitfall outlined by~\cite{arp2022and}.
    \item \textbf{Temporal Inconsistency}\footnote{Temporal inconsistency occurs when samples created chronologically after those in the test set are included in the training set, leading to unrealistic forward-looking performance.}: We partition malware samples between training and test sets based on their discovery date (see \refsec{subsubsec:data-coll}), preventing the model from inadvertently learning about future threats during training.
\end{itemize}
This rigorous data processing framework establishes a solid foundation for training and fairly evaluating our baseline detectors.

\begin{table}[t!]
    \centering
    \small
    \begin{tabular}{lll}
        \toprule
        \textbf{\#} & \textbf{Events} &\textbf{Description} \\
        \midrule
        \textcolor{fs1}{1} & \{1, 2, 3, 4\} & Top 4 events from prototypes \\
        \textcolor{fs2}{2} & \{1, 2, 3, 4, 5, 6\} & Top 6 events from prototypes \\
        \textcolor{fs3}{3} & \{5, 11\} & Branch-based events \\
        \textcolor{fs4}{4} & \{5, 6, 7, 11\} & Events used in~\cite{sayadi2018customized} \\
        \textcolor{fs5}{5} & \{5, 14, 17, 21\} & Events used in~\cite{sayadi2018ensemble} \\
        \textcolor{fs6}{6} & \{1, 2, 5, 19\} & Events used in~\cite{makrani2022accelerated} \\
        \textcolor{fs7}{7} & \{2, 8, 13, 14\} & Data cache and TLB events \\
        \textcolor{fs8}{8} & \{5, 10, 11, 15\} & Branch and instruction cache events \\
        \textcolor{fs9}{9} & \{5, 7, 11, 12\} & Branch and cache events \\
        \textcolor{fs10}{10} & \{1, 3, 4, 5\} & Branch and memory events \\
        \textcolor{fs11}{11} & \{4, 5, 6, 11\} & Events used in~\cite{bahador2014hpcmalhunter} \\
        \bottomrule
    \end{tabular}
    \caption{\textbf{Feature sets} used in this work.}
    \label{tab:feature-sets}
\end{table}

\begin{table}[t!]
    \centering
    \small
    \begin{tabular}{llp{2.2in}}
        \toprule
        \textbf{\#} & \textbf{Name} & \textbf{Description} \\
        \midrule
        1 & \dt & Decision tree to replicate~\cite{he2021machine} \\
        2 & \rf & Random forest to replicate~\cite{he2021machine} \\
        3 & \svd & SVD transformation with RF to replicate~\cite{bahador2014hpcmalhunter} \\
        4 & \reptree & RepTree to replicate~\cite{sayadi2018ensemble} \\
        5 & \mlp & Multi-Layer Perceptron to replicate~\cite{sayadi2018ensemble,sayadi20192smart} \\
        \bottomrule
    \end{tabular}
    \caption{\textbf{ML-Based detection models}.}
    \label{tab:models}
\end{table}

\begin{table}[t!]
    \centering
    \small
    \begin{tabular}{llllll}
        \toprule
        Model & Accuracy & F1 & Recall & Precision & FPR \\
        \midrule
        \ens(1) & 0.878 & 0.615 & 0.993 & 0.446 & 0.134 \\
        \ens(2) & 0.800 & 0.494 & 0.998 & 0.328 & 0.222 \\
        \ens(3) & \textbf{0.928} & \textbf{0.731} & 0.995 & \textbf{0.578} & \textbf{0.079} \\
        \ens(4) & 0.863 & 0.588 & 0.998 & 0.417 & 0.152 \\
        \ens(5) & 0.886 & 0.630 & 0.989 & 0.462 & 0.125 \\
        \ens(6) & 0.833 & 0.539 & \textbf{0.999} & 0.369 & 0.185 \\
        \ens(7) & 0.889 & 0.637 & 0.993 & 0.469 & 0.122 \\
        \ens(8) & 0.919 & 0.706 & 0.995 & 0.547 & 0.090 \\
        \ens(9) & 0.874 & 0.607 & 0.998 & 0.437 & 0.140 \\
        \ens(10) & 0.796 & 0.489 & 0.996 & 0.324 & 0.225 \\
        \ens(11) & 0.853 & 0.570 & 0.997 & 0.400 & 0.163 \\
        \bottomrule
    \end{tabular}
    \caption{\textbf{Performance of ensemble detectors}.}
    \label{tab:baseline-perf-1-slice-ens}
\end{table}

\subsection{Selecting Relevant Hardware Events} \label{subsec:feat-sel}

\begin{figure*}[t!]
    \centering
    \includegraphics[width=0.95\textwidth]{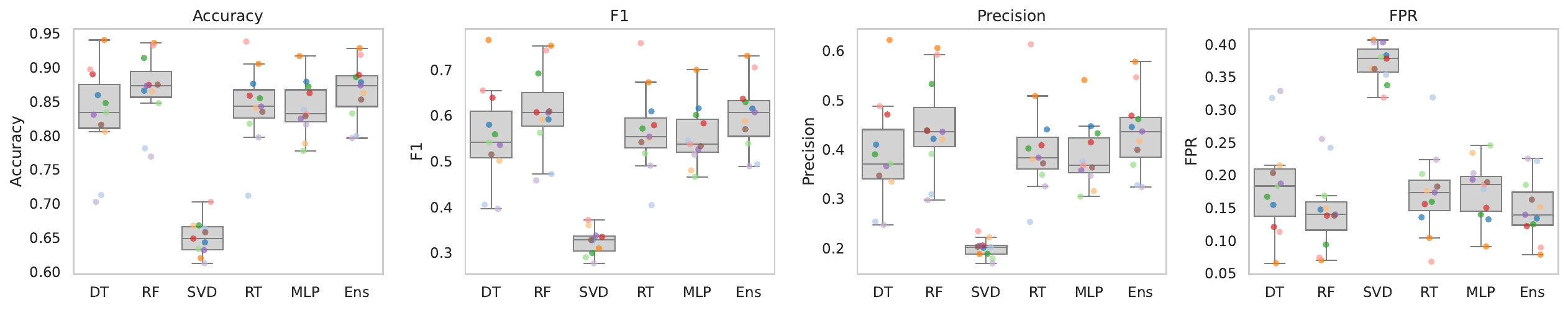}
    \caption{\textbf{Performance of individual models}. Each plot shows the performance distribution of a given model family (e.g., \dt, \rf) for a specific metric. Points are color-coded by feature set across all plots. The box denotes the interquartile range (middle 50\%), while the whiskers indicate the non-outlier range. For the rightmost plot (FPR), lower values indicate better performance.}
    \label{fig:ind-models-boxplots-1-slice-color}
\end{figure*}

The selection of which hardware events to monitor is a critical design choice, constrained by the finite number of HPCs available on commercial processors. To maximize the discriminative power of our models under this constraint, we identify relevant features through two complementary approaches: (1) applying a state-of-the-art multivariate time-series (MTS) feature selection technique, and (2) leveraging events consistently highlighted as informative in prior literature.

Our starting point is the set of events listed in \reftab{tab:features}, which encompasses events commonly used in related research~\cite{malone2011hardware, sayadi2018ensemble, he2021machine, zhou2018hardware, demme2013feasibility, basu2019theoretical, bahador2014hpcmalhunter}.

For the first approach, we implement the scalable, classifier-agnostic MTS feature selection method proposed by~\cite{dhariyal2023scalable}. This technique constructs a per-class \emph{prototype} for each hardware event by aggregating (e.g., via mean or median) its trace across all samples within a class. The relevance of an event is then quantified by the Euclidean distance between the malware and benignware prototypes for that event. A larger inter-class distance indicates greater discriminative potential. This method is specifically designed for MTS data, is computationally efficient, and is robust to noise, offering advantages over filter or wrapper methods used in earlier studies. The events in \reftab{tab:features} are ranked according to this distance metric.

For the second approach, we directly adopt the top-ranked event sets reported in key prior studies~\cite{sayadi2018customized, sayadi2018ensemble}.
Furthermore, we manually compose supplementary event sets based on recurring themes in the literature. These include sets focused exclusively on branch prediction behavior (e.g., \texttt{branch-loads}, \texttt{branch-misses}), sets combining cache and branch events, and others that blend events from different microarchitectural components. \reftab{tab:feature-sets} enumerates the specific sets of hardware events (referenced by their IDs from \reftab{tab:features}) used throughout our experiments. This multi-pronged selection strategy ensures our baselines are informed by both data-driven relevance and established domain knowledge.

\subsection{Performance of Baseline Models} \label{subsec:baseline-model-perf}

To establish baselines, we deploy several ML classifier architectures that have been used in prior hardware-assisted malware detection studies. The specific models, along with their sources, are listed in \reftab{tab:models}. We focus on architectures that are relatively simple and offer fast inference while still achieving strong performance as reported in the literature. This choice is made without loss of generality: as discussed in \refsec{sec:mechanism}, \mech can be extended to arbitrary sets of model architectures.

In addition to individual models, we employ ensemble methods to enhance detection robustness, a strategy widely adopted in related research~\cite{sayadi20192smart, sayadi2018ensemble, he2021machine}. Ensemble methods aggregate predictions from multiple base models to produce a final prediction that is often more accurate and stable than any single constituent. While techniques like boosting and bagging vary in their aggregation mechanics, their shared objective is to improve overall performance. In this work, we construct ensemble baselines using a straightforward yet effective majority voting scheme across all model instances trained on the \emph{same} feature set, similar to the approach in~\cite{khasawneh2018ensemblehmd}. We denote an ensemble using feature set $i$ as \ens($i$).

The performance of these ensemble baselines is summarized in \reftab{tab:baseline-perf-1-slice-ens}, reporting standard metrics: accuracy, F1 score, recall, precision, and false positive rate (FPR). The top-performing baseline, \ens(3), achieves an accuracy of 92.8\%, an F1 score of 0.731, and an FPR of 0.079. A notable pattern across all baselines is high recall coupled with relatively low precision, which depresses the F1 score. This occurs because, in our 90\%-benign test set, even a moderately low FPR generates a substantial number of false positives relative to the small pool of true positives (malicious samples), thereby reducing precision.

\reffig{fig:ind-models-boxplots-1-slice-color} provides further insight by illustrating the performance distributions of individual models and their corresponding ensembles. Overall, most model families operate within comparable performance ranges, with the exception of the SVD-based model. Nevertheless, prior work has shown that ensembles incorporating weaker classifiers can still achieve strong aggregate performance~\cite{sayadi2018ensemble, khasawneh2018ensemblehmd}.

We also observe that certain feature sets consistently rank among the top performers, while others persistently lag behind (colors correspond to event set indices in \reftab{tab:feature-sets}). For example, feature sets 2 and 10 (pastel blue and purple) frequently appear near the bottom, whereas feature sets 3 and 8 (darker orange and pink) consistently achieve the highest performance. This indicates that some feature sets possess substantially greater predictive power in isolation, largely independent of the underlying model family. Finally, while ensemble baselines do not always surpass the best individual models, they generally perform on par with the top performer, offering robust overall performance.

\begin{figure}[t!]
    \centering
    \includegraphics[width=0.95\columnwidth]{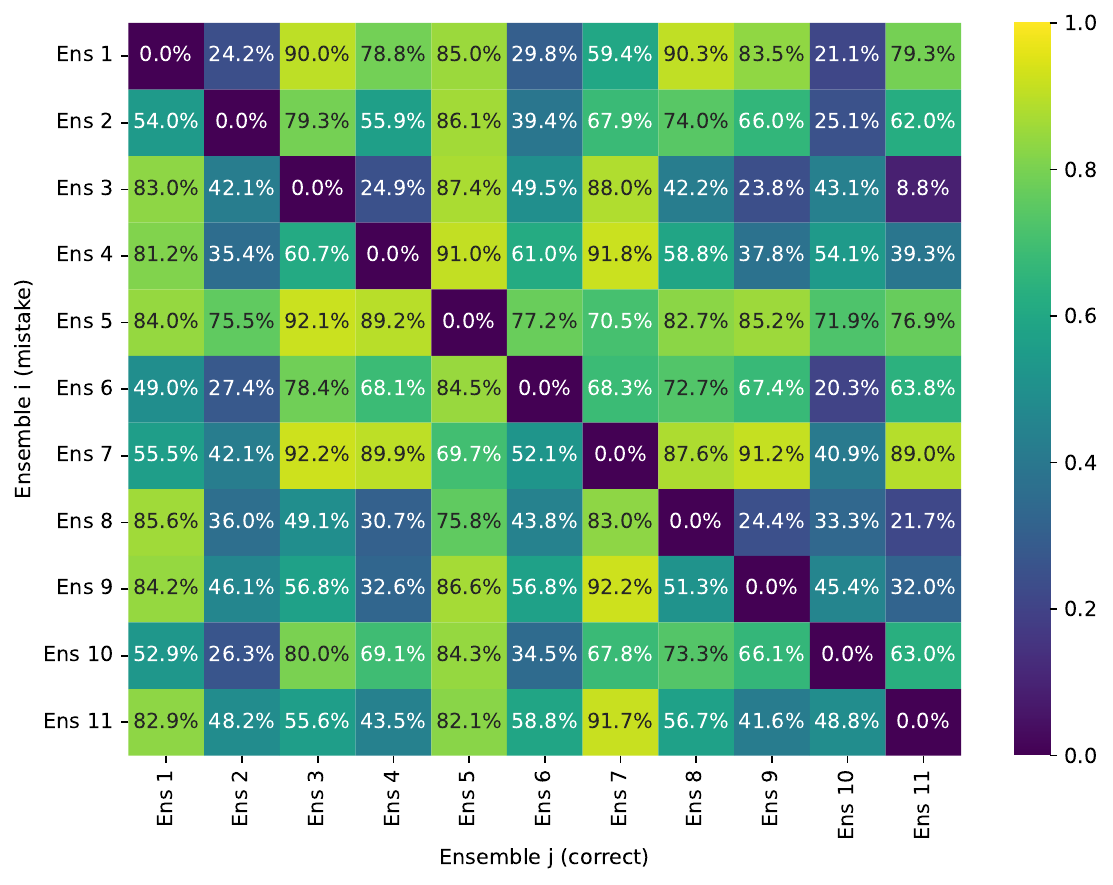}
    \caption{\textbf{Percentage of caught mistakes.} The heatmap illustrates mistake coverage across ensemble baselines. For each row–column pair $(i, j)$, the value indicates the percentage of mistakes made by baseline $i$ that are correctly classified by baseline $j$. The diagonal entries are zero, since a baseline cannot, by definition, correct its own errors.}
    \label{fig:mistakes-heatmap}
\end{figure}

Although the ensemble baselines achieve strong accuracy, they expose a clear opportunity to improve the balance between precision and recall, as reflected in their F1 scores and false positive rates (FPRs). For context, a naive classifier that predicts \emph{every} sample as benign would attain 90\% accuracy and 0\% FPR on our test set, yet would exhibit zero recall and an undefined F1 score. Our objective, therefore, is not merely to increase accuracy, but to meaningfully improve F1 and FPR by strategically combining the strengths of multiple baselines trained on complementary feature sets.

To highlight the benefit of incorporating multiple feature sets, for each ensemble model we compute the percentage of its mistakes that are correctly classified by the other baselines, as shown in \reffig{fig:mistakes-heatmap}. As illustrated, different models excel at correcting the errors made by others. For instance, \ens(8) recovers the highest proportion of mistakes made by \ens(1), whereas for errors made by \ens(2), \ens(5) performs best. These complementary error patterns motivate the inclusion of diverse feature sets. Indeed, as we hypothesize here and demonstrate empirically in \refsec{sec:eval}, leveraging such diversity can lead to meaningful performance improvements.
\section{\mech}\label{sec:mechanism}

In the previous section, we have shown that different model architectures, and their ensemble models, exhibit varying detection performance depending on the selected feature set. Conventional malware detection pipelines typically select and deploy the single best-performing model (e.g., the one with the highest accuracy or F1-score) across the considered feature sets. This deployed model could itself be an ensemble of multiple architectures. Crucially, under this conventional approach, the selected feature set (i.e., the specific set of hardware performance events to monitor) remains fixed at all times. In contrast, the core idea of our work is to improve detection robustness by deliberately \emph{varying} the feature set over time. This section formalizes this concept and describes our method for learning an optimal schedule of feature sets.

\subsection{Problem Statement}\label{subsec:problem}

To operationalize the use of multiple feature sets, we partition the execution trace of a program into discrete, contiguous time intervals, which we refer to as \emph{slices}. We then allow a different feature set (and thus a correspondingly trained detector) to be selected for each slice. The final label for the entire execution (i.e., malicious or benign) is determined by aggregating the per-slice detection outputs, for instance, via a majority vote. We note a key hardware-imposed constraint: it is infeasible to monitor multiple distinct sets of hardware events concurrently within a single time slice. Therefore, if feature set $k$ is selected for monitoring during a given slice, \emph{only} the detector models trained on that specific feature set $k$ are applicable for classification during that slice. Detectors trained on other feature sets cannot be utilized, as the requisite input data is unavailable, and re-executing the program to collect alternate event traces is impractical.

Given this constraint and a fixed number of slices, our primary objective is to select the optimal sequence of detectors across slices. Formally, a \emph{detector} is a model (e.g., a decision tree, random forest, or neural network) trained on a specific feature set. In this work, without loss of generality, we focus on detectors that are themselves ensemble models, as discussed in \refsec{subsec:baseline-model-perf}; our method can be trivially extended to sequences of individual classifiers. Similarly, we present our framework using three time slices and a majority-vote aggregation rule for the final label. Our methodology generalizes naturally to other numbers of slices and alternative aggregation methods (e.g., weighted voting or threshold-based rules).

Building upon the methodology described in \refsec{sec:methodology}, we assume a set of eleven distinct ensemble detectors, denoted \ens($i$) for $i \in \{1, \dots, 11\}$, each trained on one of the eleven candidate feature sets. For a three-slice execution, the space of possible detector sequences $S$ contains $11^3 = 1331$ distinct combinations (e.g., one sequence might be $[\ens(10), \ens(2), \ens(4)]$, while another is $[\ens(1), \ens(4), \ens(11)]$). Our goal is to identify the sequence that yields the most robust and accurate overall detection.

\subsection{Formal Model}\label{subsec:formal}

To identify the optimal detector sequence, we adopt a data-driven learning approach. We split the available training data into two disjoint subsets. First, 80\% of the data is used to train the base classifiers and construct the eleven ensemble detectors, one per feature set. The remaining 20\% of the training data, denoted as $\mathcal{I}$, is reserved to learn the optimal sequence of detectors across slices. While using only 80\% of the data for initial detector training could potentially reduce their individual performance, our hypothesis (validated in \refsec{sec:eval}) is that the information gained from strategically scheduling detectors using $\mathcal{I}$ leads to superior overall detection performance compared to any fixed, single-feature-set deployment that uses all the data for training.

We model the problem of learning the best sequence as an optimization problem. Let $S$ be the set of all $11^3$ possible sequences. We introduce a decision variable $x_s$ for each sequence $s \in S$, representing the probability that sequence $s$ is selected for deployment. For each training sample $i \in \mathcal{I}$ and each sequence $s$, we define $p_{i,s}$ as the prediction confidence that sequence $s$ assigns to sample $i$ being malware. In its simplest form, $p_{i,s}$ could be a binary $0$ or $1$ (benign or malicious). However, we can provide richer information to the learning algorithm by letting $p_{i,s} \in [0, 1]$ represent a calibrated confidence score. The method for aggregating per-model confidence scores to obtain a meaningful $p_{i,s}$ for an entire sequence is detailed in Subsection~\ref{subsubsec:prediction}.

Let $y_i \in \{0, 1\}$ denote the true label of sample $i$ (with $1$ indicating malware). The sequence learning process can be formulated as the following constrained optimization problem:
\begin{align}
    \label{equ:opt}
    \begin{aligned}
        & \underset{x}{\text{maximize}}
        & & F(x) \\
        & \text{subject to}
        & & \sum_{s \in S} x_s = 1, \quad x_s \ge 0 \quad \forall s \in S,
    \end{aligned}
\end{align}

where $F(x)$ is an objective function that measures the quality of the probabilistic mixture of sequences defined by $x$, given the predictions $p_{i,s}$ and true labels $y_i$. There are multiple choices for $F(x)$, such as the negative mean squared error or the negative mean absolute error. The specific form of $F(x)$ used in \mech is introduced in \refsec{subsubsec:objective}.

\subsubsection{Aggregated Predictions}\label{subsubsec:prediction}

To define the sequence-confidence vector $p_{i,s}$ for sample $i$ under sequence $s$, we require a principled method for aggregating the probabilistic outputs of the individual classifiers within each ensemble detector in $s$, and subsequently aggregating the outputs of the ensemble detectors that constitute $s$. Suppose for a given slice, the assigned ensemble detector comprises $K$ base models. Let each model $k$ output a posterior probability $q_k = \Pr(Y=1 \mid i, \mathcal{M}_k)$, where $\mathcal{M}_k$ represent the output of model $k$. A naive average of these $q_k$ values lacks a rigorous probabilistic interpretation, as probabilities are not additive measures of evidence.

Instead, we transform $q_k$ to the log-odds (logit) scale:
\[
    \ell_k \triangleq \log \frac{q_k}{1 - q_k} = \log \frac{\pi}{1-\pi} + \log \frac{\Pr(\mathcal{M}_k \mid i, Y=1)}{\Pr(\mathcal{M}_k \mid i, Y=0)},
\]
where $\pi$ is the prior probability of class $1$.
This transformation maps the prediction to a real-valued quantity interpretable as a log-likelihood ratio. Positive $\ell_k$ indicates evidence for class $1$ (malware), negative $\ell_k$ for class $0$ (benign), with magnitude representing strength of evidence. Under the assumption that models provide conditionally independent evidence given the true label, Bayes' rule implies that:
\[
    \Pr(Y=1 \mid i, \mathcal{M}_{1:K}) \propto \prod_{k=1}^K \Pr(\mathcal{M}_k \mid i, Y=1) \cdot \Pr(Y = 1).
\]
Therefore, the combined log-odds are additive:
\[
\log \frac{\Pr(Y=1 \mid i, \mathcal{M}_{1:K})}{\Pr(Y=0 \mid i, \mathcal{M}_{1:K})} = b + \sum_{k=1}^K \ell_k,
\]
where $b = -(K - 1) \pi / (1 - \pi)$ is a constant. In contrast to averaging raw probabilities, which forms a linear opinion pool and tends to dilute strong evidence, aggregating in logit space preserves likelihood ratios and naturally accumulates independent signals.

Inspired by this, for each time slice within sequence $s$, we compute the average%
\footnote{Summing logits corresponds to accumulating independent log-likelihood ratios, whereas averaging logits normalizes the aggregate evidence and prevents confidence from growing with ensemble size. This provides a more practical regularization when base classifiers are correlated or imperfectly calibrated.}
log-odds across its assigned ensemble. We then obtain a slice-level confidence via the sigmoid function: $\sigma(1/K \sum_{k} \ell_k)$. Averaging log-odds combines models in evidence space, and the subsequent sigmoid maps this aggregated evidence back to a calibrated probability. To obtain the final sequence-level confidence $p_{i,s}$ for sample $i$, we average the slice-level log-odds and again apply the sigmoid. This aggregation in log-odds space emphasizes predictions from confident models while attenuating those from uncertain ones, yielding a principled and well-calibrated measure of sequence confidence.

\subsubsection{Objective Function}\label{subsubsec:objective}

Our setting can be viewed as an expert aggregation problem. Formally, we are given a dataset $\{(p_i, y_i)\}_{i \in \mathcal{I}}$, where $p_i = (p_{i,1}, \dots, p_{i,|S|})$ is the vector of sequence-specific prediction confidences for sample $i$. Each input dimension corresponds to a distinct expert (i.e., a sequence $s$), and the value along that dimension represents the expert's confidence. Our goal is to combine these heterogeneous probabilistic predictions into a single calibrated decision. We adopt a discriminative aggregation model based on logistic regression:
\[
    P(Y=1 \mid P = p_{i}) = \sigma\!\left(\sum_s x_s p_{i,s} \right),
\]
where $\sigma(\cdot)$ denotes the sigmoid function and $x_s$ are learnable aggregation weights. In standard logistic regression, the coefficients $x_s$ are unconstrained. However, in our setting, to obtain interpretable aggregation weights, we restrict $x$ to lie on the probability simplex. This can be viewed as a calibrated linear opinion pool: the simplex-constrained weights define a global mixture over experts, while the outer sigmoid learns a nonlinear calibration that maps the aggregated confidence to a final posterior probability. Under this model, the posterior log-odds of the malware class given the sequence predictions is modeled linearly:
\[
    \log\frac{\Pr(Y = 1 \mid P = p_i)}{\Pr(Y = 0 \mid P = p_i)} = x^T p_i.
\]

The parameter vector $x$ (defining the mixture of sequences) is then fit by maximizing the log-likelihood of the observed data $\mathcal{I}$, plus a regularization term $\lambda R(x)$ to prevent overfitting ($\lambda \ge 0$). This yields the objective function:
\[
    F(x) \;\triangleq\; \sum_{i \in \mathcal{I}} \Bigl( y_i (x^T p_i) - \log\bigl(1 + \exp(x^T p_i)\bigr) \Bigr) - \lambda R(x).
\]

Maximizing $F(x)$ corresponds to performing logistic regression on the sequence-confidence vectors $p_i$, where the coefficients $x$ directly encode the importance (or selection probability) of each sequence. This perspective provides a clear intuition for our modeling choice. Logistic regression offers a principled, likelihood-based mechanism for learning how individual expert confidences translate into a final binary decision. Under this formulation, the learned weights can be interpreted as global confidence scores over experts, estimated discriminatively by maximizing the log-likelihood of the aggregated confidences.

\subsection{Putting It All Together} \label{subsec:pipeline}

Having established the individual components of \mech, we now synthesize these elements into a cohesive workflow. This subsection outlines the complete pipeline for \mech, detailing both the offline training procedure and the online deployment for real-time malware detection. The integrated process is formalized in \refalg{alg:hydra}.

The pipeline operates in two distinct phases. The first phase is performed \emph{offline}, where the system is trained and configured using a labeled dataset. The second phase is executed \emph{online} during system deployment, where the trained mechanism analyzes new, unseen executable samples.

\begin{algorithm}[t!]
    \begin{algorithmic}[1]
        \Require Set of feature sets $F$, set of model architectures $K$, labeled training data $T$
        \LComment{Sequence Training (Offline)}
        \State Split $T$ into $M$ \& $I$ for model and sequence training
        \State Train classifiers on $M$ for all $f \in F, k \in K$
        \State Construct set of ordered sequences of detectors $S$
        \State Aggregate predictions $p_{s,i}$ for all $s \in S, i \in I$ (\refsec{subsubsec:prediction})
        \State Find optimal sequence probabilites $x_s$ using \refequ{equ:opt}
        \LComment{Malware Detection (Online)}
        \State Draw sample sequence according to $x_s$
        \State Track events according to selected sequence
        \State Deploy corresponding detectors for malware detection
    \end{algorithmic}
\caption{\mech}
\label{alg:hydra}
\end{algorithm}

The offline phase (Lines 2–6) is computationally intensive but performed only once. The partitioning prevents data leakage and ensures that sequence optimization is evaluated on data unseen by the base classifiers. Model training defines the space of possible execution strategies: for $|F|$ feature sets and $|K|$ model architectures, the total number of sequences of length~$L$ is $(|F|\times|K|)^L$. In practice, this space can be managed by filtering sequences based on preliminary validation performance or by employing heuristic search. In our experimental methodology, we significantly reduce this space to $|F|^L$ by focusing exclusively on ensemble models. The core of \mech is the resulting optimization problem, which learns a probability distribution over sequences that maximizes expected detection utility.

The online phase (Lines 8–10) is designed for low-latency, real-time operation. The stochastic selection of sequences ensures non-deterministic system behavior, which can complicate evasion attempts targeting a fixed monitoring configuration. Monitoring events is subject to the practical constraint that only the limited set of hardware counters required by the first classifier in the selected sequence can be observed simultaneously; subsequent classifiers may require different events, necessitating reconfiguration of performance counters. The final step produces the classification decision by aggregating the outputs of the sequence’s constituent detectors.

This integrated pipeline embodies the key innovations of \mech: it moves beyond static, single-feature-set detection by selecting and sequencing multiple detectors. By optimizing the sequence distribution offline, the system balances detection performance against resource costs. \refsec{sec:eval} empirically evaluates this pipeline, measuring its effectiveness against established baselines and across diverse threat models.

\section{Evaluation} \label{sec:eval}

\subsection{Experimental Setup} \label{subsec:exp-setup}

\subsubsection{Execution Environment} \label{subsubsec:exec-env}

We conduct our experiments on an AMD Ryzen Threadripper 3945WX processor, which provides six core performance event counters per core, enabling the simultaneous monitoring of up to six distinct per-thread hardware events. Although additional counters are available (six for L3 complex events and four for data fabric events), we do not utilize them. Events associated with shared resources, such as L3 cache misses, can exhibit high runtime variance due to interference from co-executed workloads. To ensure consistency, we therefore focus exclusively on private per-thread counters and their associated events. We refer readers to \refsec{subsec:feat-sel} for a discussion of event selection.

Our testbed runs Ubuntu 20.04 LTS. Each binary is executed within a Linux container using \texttt{lxc/incus}~\cite{incus} (version~6.0.3), and hardware event data are collected using \texttt{perf}~\cite{Linux} (version~5.15.168). For data collection, we configure \texttt{perf} to sample hardware events at 10ms intervals. The testbed system is equipped with 16,GB of DDR4 memory operating at 3200,MT/s and NVMe storage.

We implement the sequence training phase of \mech using \texttt{CVXPY}~\cite{diamond2016cvxpy}. Unless stated otherwise, we use the negative logistic loss as the objective function (see \refsec{subsubsec:objective}), employ mean logit aggregation to combine ensemble- and slice-level confidences (see \refsec{subsubsec:agg-methods}), and set $\lambda = 0$ (no regularization term).

\subsubsection{Collecting Traces} \label{subsubsec:data-coll}

We collect HPC traces, which capture hardware event activity, from a diverse set of benign and malicious binaries. Our dataset includes approximately 5{,}000 malware binaries obtained from the \texttt{MalwareBazaar} repository~\cite{MalwareBazaar}, spanning a range of families, including trojans, droppers, downloaders, viruses, and backdoors. Because new malware samples are periodically released to the repository, we collect binaries over time and tag them according to their release timestamps, as recorded in the metadata provided by \texttt{MalwareBazaar}. These temporal tags are subsequently used to partition the malware samples into training and test sets (see \refsec{subsec:data-proc} for details).

To collect each malware trace, we instantiate a fresh container prepopulated with benign files to support malware samples that interact with the local filesystem. Because only six per-thread HPCs are available, each binary is executed four times to capture all 21 events. We discard any execution in which the malware fails to run for at least a predefined duration (10~seconds in our experiments). For \mech, we divide each sample further into three equal-length slices (see \refsec{subsubsec:prediction}).

\begin{table}[t!]
    \centering
    \small
    \begin{tabular}{lp{4cm}c}
        \toprule
        Suite & Workloads & Source\\
        \midrule
        Byte-UnixBench & All workloads & \cite{unixbench}\\
        ClickBench & \texttt{cockroachdb}, \texttt{cratedb}, \texttt{sqlite} & \cite{clickbench}\\
        CortexSuite & \texttt{word2vec} & \cite{cortexsuite}\\
        DBENCH & All workloads & \cite{dbench}\\
        LMBench & All workloads & \cite{mcvoy1996lmbench}\\
        Memcached & N/A & \cite{memcached}\\
        Memtier & N/A & \cite{memtier}\\
        MiBench & All workloads & \cite{guthaus2001mibench}\\
        PARSEC & All workloads & \cite{Bienia2008PARSEC} \\
        Splash & All workloads& \cite{sakalis2016splash}\\
        Phoronix & All workloads & \cite{phoronix}\\
        Redis & N/A & \cite{redis}\\
        Apache Spark & \texttt{kmeans}, \texttt{trans-closure}, \texttt{pagerank}, \texttt{tpcds} & \cite{zaharia2016apache} \\
        SPEC 2017 & All workloads & \cite{bucek2018spec} \\
        Stress-ng & \texttt{af-alg}, \texttt{cpu}, \texttt{crypt}, \texttt{device}, \texttt{iomix}, \texttt{locks}, \texttt{matrix}, \texttt{mq}, \texttt{pipes}, \texttt{sock} & \cite{stress-ng}\\
        Sysbench & All workloads & \cite{sysbench}\\
        Python scripts & In-house ML scripts & N/A \\
        \bottomrule
    \end{tabular}
    \caption{\textbf{Benchmarks} used in this work.}
    \label{tab:benchmarks-used}
\end{table}

To collect benignware samples, we source binaries from a variety of benchmark suites (see \reftab{tab:benchmarks-used} for a complete list). A subset of these benchmarks, including \texttt{SPEC CPU 2017}, \texttt{MiBench}, \texttt{LMBench}, and \texttt{UnixBench}, has been used in prior hardware-based malware detection studies~\cite{he2021machine, sayadi20192smart, kadiyala2020hardware, demme2013feasibility}.
We segment benign execution traces into disjoint individual samples.\footnote{Splitting longer benignware traces into smaller samples prevents detection models from relying on behaviors specific to the start of benign programs. Because benign workloads often run for extended periods, segmentation exposes models to diverse execution phases rather than biasing them toward initial program states.} In total, we collect approximately 19{,}000 benignware samples. Although each benchmark is split into multiple samples, we treat each benchmark as an indivisible unit when partitioning the data, thereby preventing leakage across model training, sequence training, and test sets.

\subsection{Evaluation Metrics} \label{subsec:metrics}

We evaluate all baselines and \mech using accuracy, F1, precision, recall, and false positive rate. Accuracy, defined as the proportion of correct predictions, is a standard performance metric across many domains; however, it can be misleading in the presence of class imbalance. Because our test set contains 90\% benign samples (see \refsec{subsec:data-proc}), a trivial classifier that predicts every sample as benign would already achieve 90\% accuracy. We therefore also report precision and recall, along with their harmonic mean, F1.

In the binary setting, precision ($p$) measures the fraction of samples predicted as malware that are truly malicious, while recall ($r$) measures the fraction of malicious samples that are correctly identified. The F1 score combines these metrics as $2pr/(p+r)$. F1 penalizes models that achieve high recall at the expense of low precision (i.e., many false alarms), as well as models with high precision but low recall. These metrics are commonly used in related work~\cite{cheng2023feasibility,sayadi20192smart}.

\subsection{\mech's Performance} \label{subsec:mech-perf}

\begin{table}[t!]
    \centering
    \small
    \begin{tabular}{llllll}
        \toprule
        Model & Accuracy & F1 & Recall & Precision & FPR \\
        \midrule
        \mech & 0.971 & 0.872 & 0.997 & 0.775 & 0.031 \\
        \midrule
        \ens(1) & 10.59 & 41.74 & 0.36 & 73.93 & 76.61 \\
        \ens(2) & 21.49 & 76.67 & -0.06 & 136.34 & 85.87 \\
        \ens(3) & \textbf{4.64} & \textbf{19.32} & 0.24 & \textbf{34.15} & \textbf{60.23} \\
        \ens(4) & 12.56 & 48.35 & -0.06 & 86.00 & 79.30 \\
        \ens(5) & 9.64 & 38.54 & \textbf{0.79} & 67.90 & 74.95 \\
        \ens(6) & 16.67 & 61.83 & \textbf{-0.18} & 110.06 & 83.08 \\
        \ens(7) & 9.26 & 36.96 & 0.36 & 65.41 & 74.35 \\
        \ens(8) & 5.75 & 23.63 & 0.18 & 41.87 & 65.02 \\
        \ens(9) & 11.19 & 43.61 & -0.06 & 77.57 & 77.56 \\
        \ens(10) & \textbf{21.99} & \textbf{78.25} & 0.06 & \textbf{139.05} & \textbf{86.09} \\
        \ens(11) & 13.90 & 52.92 & -0.00 & 94.08 & 80.73 \\
        \bottomrule
    \end{tabular}
    \caption{\textbf{Performance} of \mech and relative improvements compared to \ens baselines.}
    \label{tab:lp-baseline-percent}
\end{table}

\begin{figure*}[t!]
    \centering
    \includegraphics[width=0.95\textwidth]{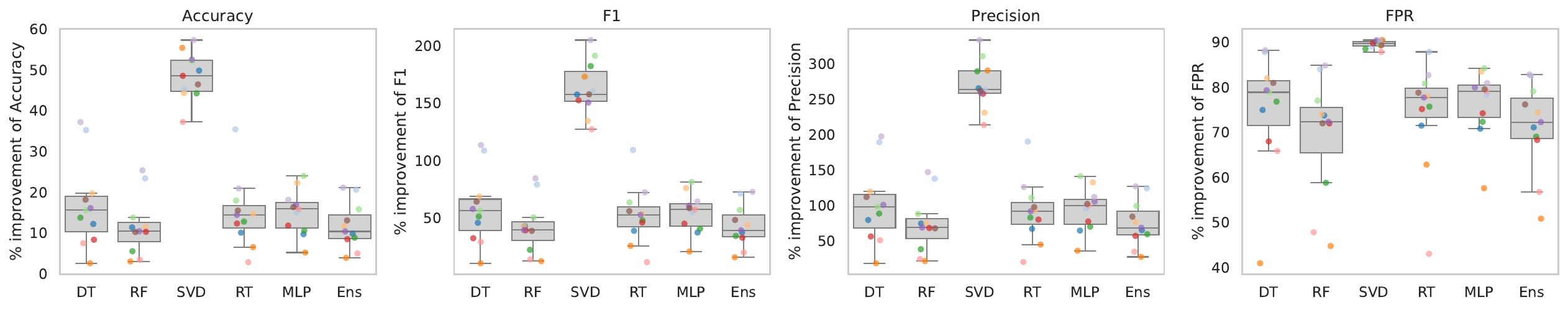}
    \caption{\textbf{Percent-improvement} boxplots of \mech using individual baselines relative to all baselines. Values above zero indicate improved performance. For FPR (far right), positive values indicate a reduction in false positive rate achieved by \mech relative to the baselines.}
    \label{fig:sensitivity-all-models}
\end{figure*}

To evaluate the performance of \mech, we split the full dataset into a training set of approximately 7{,}000 samples and a test set of approximately 17{,}000 samples, with the test set comprising 90\% benign and 10\% malicious samples. For the baselines, we use the entire training set to train $5\times 11 = 55$ classifiers, corresponding to every combination of feature set (\reftab{tab:feature-sets}) and model architecture (\reftab{tab:models}). We then construct an ensemble model \ens($i$) for each feature set $i$ using simple majority voting. These baselines represent state-of-the-art hardware-based malware detection approaches, as discussed in \refsec{sec:background}.

\mech's sequence learning in \refequ{equ:opt} produces a selection distribution over all sequences. To evaluate \mech's performace, we compute the expected performance on the test set for each metric $m$ as follows:
\[
    \mathbb{E}_{s \sim x_s}\!\left[m(s)\right] \; = \; \sum_{s \in S} x_s\, m(s).
\]

\reftab{tab:lp-baseline-percent} reports the performance of \mech across multiple detection metrics in the top row, followed by the percentage improvements of \mech relative to the \ens baselines in the remaining rows, with the largest and smallest gains for each metric highlighted in bold. As shown, \mech achieves an accuracy of 0.971 and an F1 score of 0.872, while maintaining a false positive rate (FPR) of only 3.1\%. \mech outperforms all individual baselines. Even when compared to \ens(3), the best-performing baseline across the evaluated metrics, \mech attains a 4.64\% improvement in accuracy, a 19.32\% improvement in F1 score, and a 60.23\% decrease in FPR.

Ensemble baselines are robust, and using them as baseline detectors for \mech dramatically reduces the size of the sequence space. However, they incur higher inference-time costs (see \refsec{subsec:cost} for the discussion on costs). As such, it may be preferable to use individual models as baseline detectors in \mech. As discussed in \refsec{subsec:pipeline}, however, this expands the search space to $(|F|\times|K|)^L$; in our setup, the total number of possible sequences is $55^3$, or 166{,}375. Training sequences at this scale can impose substantial computational overhead. While this may be acceptable in some settings--since sequence training is performed offline and lies outside the critical path--it may nevertheless be infeasible in others.

The approach we adopt in this paper is to reduce this space to a subset of top-performing model architectures and feature sets. To identify strong candidates, we evaluate individual models on a portion of the sequence-training data. Based on this evaluation, we select the model architectures ${\rf, \reptree, \mlp}$ and the feature sets ${1, 3, 5, 7, 8}$, resulting in $(3 \times 5)^3 = 3375$ distinct sequences. We note that this is a heuristic strategy and may eliminate high-performing sequences, as combining individually strong models does not necessarily yield optimal sequences, and the best solutions may involve models that appear weaker in isolation.

\reffig{fig:sensitivity-all-models} shows the distribution of percent improvement of \mech over all baselines. \mech outperforms all baselines across all metrics, achieving 96.4\% accuracy, 0.845 F1, and a 3.88\% FPR. These results indicate that \mech generalizes effectively to less expensive (non-ensemble) baselines, even when restricted to a limited subset of the full sequence space.

\subsubsection{Sensitivity to Objective Function} \label{subsubsec:loss-func}

\begin{figure}[t!]
    \centering
    \includegraphics[width=0.95\columnwidth]{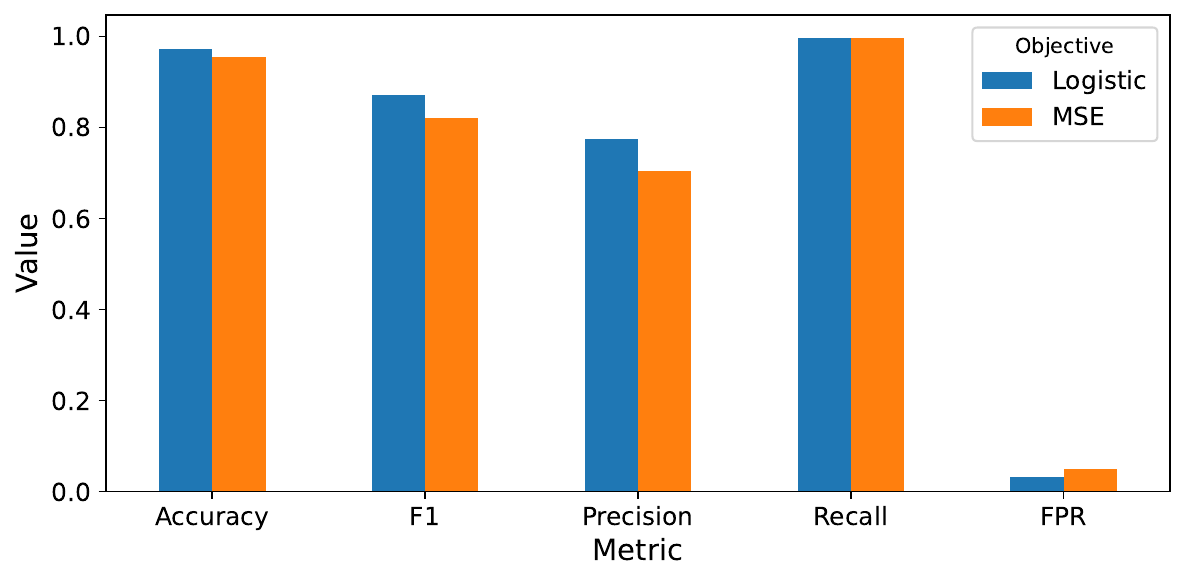}
    \caption{\textbf{Performance comparison} of \mech using the logistic and MSE objective functions. Both configurations employ mean logit probability aggregation and no regularization.}
    \label{fig:sensitivity-loss-func}
\end{figure}

In this section, we empirically evaluate the effectiveness of two functions as the objective in \refequ{equ:opt}: negative logistic loss (see \refsec{subsubsec:objective}) and negative mean squared error (MSE). The F1 score of \mech under each objective function is shown in \reffig{fig:sensitivity-loss-func}. Logistic loss consistently outperforms MSE across all tracked metrics, achieving 97.1\% accuracy, 0.872 F1, and a 3.1\% FPR, compared to 95.4\% accuracy, 0.82 F1, and 5.1\% FPR with MSE.

One reason for these results is that the logistic objective corresponds to maximizing the log-likelihood under a Bernoulli model (as discussed in \refsec{subsubsec:objective}). This yields calibrated probabilities, leads to a convex and well-behaved optimization problem, and properly penalizes confident mistakes. In contrast, MSE implicitly assumes Gaussian noise and treats probabilities as ordinary real-valued targets. Moreover, combining MSE with a sigmoid can result in poor gradient behavior and less reliable learning dynamics.

\subsubsection{Sensitivity to Aggregation Methods} \label{subsubsec:agg-methods}

Different aggregation strategies for probabilities and predictions can lead to substantially different sequence-training outcomes. In this section, we evaluate \mech under both logistic and MSE objectives using four aggregation methods: (i) average of logits followed by a sigmoid (our default configuration), (ii) sum of logits followed by a sigmoid, (iii) direct averaging of probabilities, and (iv) majority voting over hard predictions (i.e., without probabilistic outputs). We refer to these methods as \textit{logit-mean}, \textit{logit-sum}, \textit{mean}, and \textit{preds}, respectively.

As shown in \reftab{tab:sensitivity-agg-metrics}, the combination of the logistic objective with logit-mean aggregation achieves the best performance across all metrics (bolded). The second-best configuration is logistic with logit-sum aggregation, which yields an accuracy that is 1.24\% lower (95.9\%), an F1-score decrease of 5.39\% (0.825), and a 45.2\% increase in FPR (to 4.5\%).

As discussed in \refsec{subsubsec:prediction}, logit-mean aggregation acts as a form of regularization by preventing confidence values from growing with the size of the ensemble. This regularization appears to improve generalization by limiting extreme confidences, thereby allowing the loss functions to better exploit calibrated confidence information. Consistent with this intuition, mean-logit aggregation outperforms logit-sum under both logistic and MSE objectives.

Interestingly, directly averaging probabilities (mean) yields performance nearly identical to that obtained using pure predictions (preds). While one would expect confidence-aware aggregation to outperform hard voting (since binarization discards uncertainty information), the mean-probability aggregation does not appear to provide a meaningful advantage in practice. Although taking the arithmetic mean of probabilities lacks a principled probabilistic justification in this context, this empirical behavior is noteworthy.

\begin{table}[t!]
    \centering
    \footnotesize
    \begin{tabular}{llrrrrr}
        \toprule
        Loss & Method & Acc & F1 & Prec & Rec & FPR \\
        \midrule
        \multirow{4}{*}{Logistic} & logit-mean  & \textbf{0.971} & \textbf{0.872} & \textbf{0.775} & \textbf{0.997} & \textbf{0.031} \\
             & logit-sum & 0.959 & 0.825 & 0.705 & 0.993 & 0.045 \\
             & mean & 0.916 & 0.699 & 0.539 & 0.995 & 0.093 \\
             & preds & 0.915 & 0.704 & 0.549 & 0.997 & 0.094 \\
        \midrule
        \multirow{4}{*}{MSE} & logit-mean & 0.954 & 0.820 & 0.705 & 0.997 & 0.051 \\
             & logit-sum & 0.934 & 0.757 & 0.618 & 0.997 & 0.072 \\
             & mean & 0.916 & 0.699 & 0.539 & 0.995 & 0.093 \\
             & preds & 0.915 & 0.704 & 0.549 & 0.997 & 0.094 \\
        \bottomrule
    \end{tabular}
    \caption{\textbf{Performance} of \mech using different objective functions and aggregation methods.}
    \label{tab:sensitivity-agg-metrics}
\end{table}

\subsubsection{Sensitivity to Training Data} \label{subsubsec:train-data-split}

\begin{figure}[t!]
    \centering
    \includegraphics[width=0.95\columnwidth]{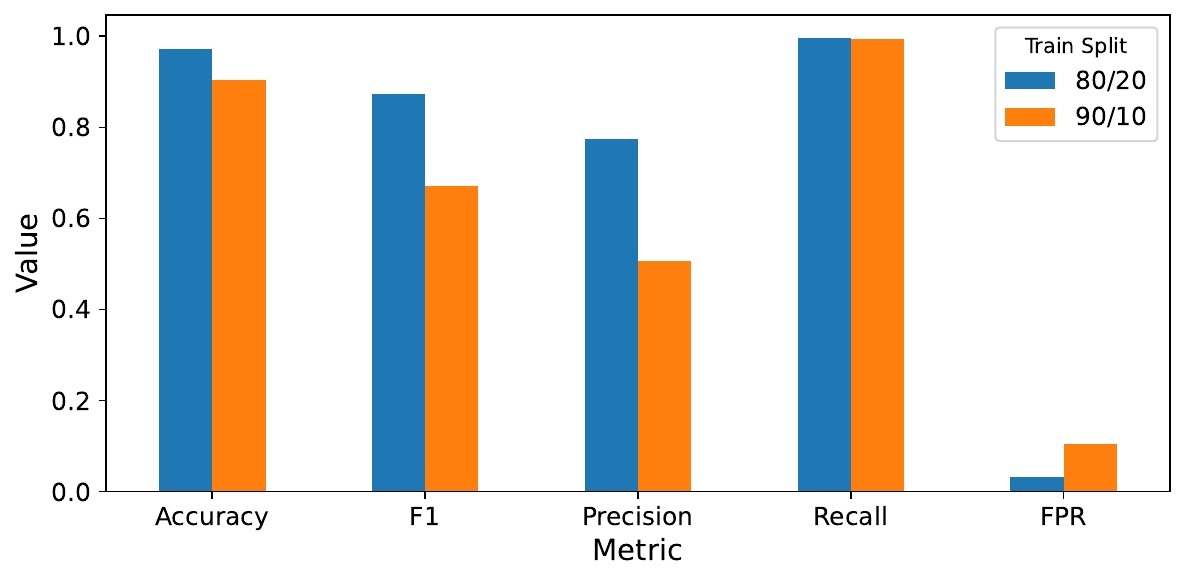}
    \caption{\textbf{Performance} of \mech when using 20\% (blue) and 10\% of the available training data for sequence training.}
    \label{fig:sensitivity-train-split}
\end{figure}

\mech splits the training data into two subsets: one for training the base models and one for sequence training. By default, \mech uses an 80/20 split between these two sets. Intuitively, allocating more data to base-model training may improve individual detector performance, but it may also degrade sequence learning due to reduced data availability. To study this trade-off, in this section, we empirically compare the default 80/20 split with a 90/10 split.

The results are shown in \reffig{fig:sensitivity-train-split}. \mech performs substantially better when 20\% of the data is reserved for sequence training compared to only 10\%. One potential explanation is that, although the base models are trained on an additional 10\% of data, this increase does not yield a meaningful improvement in their detection performance. In contrast, reducing the sequence-training set from 20\% to 10\% halves the available data and significantly diminishes its diversity. This loss of diversity arises because each benign benchmark, despite being split into multiple samples, is treated as an indivisible unit (as described in \refsec{subsubsec:data-coll}). Consequently, fewer distinct benign benchmarks are available for sequence training, further limiting \mech’s ability to learn robust and generalizable sequences.

\subsection{Inference Cost Analysis} \label{subsec:cost}

\begin{figure}[t!]
    \centering
    \includegraphics[width=0.95\columnwidth]{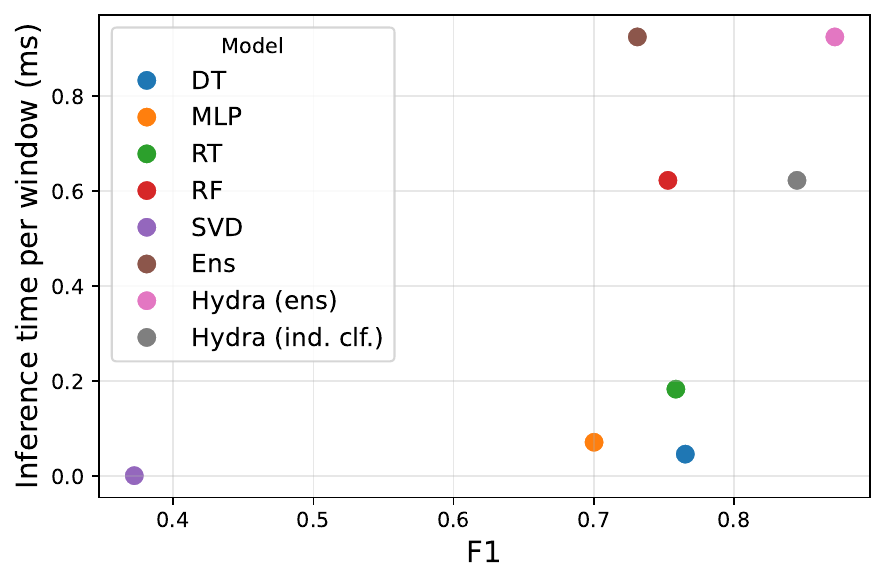}
    \caption{\textbf{F1 score versus inference time} for ensemble baselines, individual classifiers, and \mech using ensemble baselines (\mech (ens)) and individual baselines (\mech (ind. clf.)).}
    \label{fig:inf-time}
\end{figure}

Different classifiers exhibit different inference times: some are fast but less accurate, while others are slower yet more accurate, with a broad spectrum in between. \reffig{fig:inf-time} compares the F1 scores and average inference times of the baselines and \mech when using ensemble detectors. All classifiers are implemented in Python, and inference times are measured on the testbed described in \refsec{subsec:exp-setup}.

Because \mech uses ensemble baselines by default with majority-voting aggregation, it requires performing inference for each underlying classifier. To estimate inference time for \mech, we consider a worst-case scenario in which these inferences are executed sequentially.

As established earlier, \mech achieves the highest F1 score (0.872), outperforming all baselines. However, it also incurs the highest inference time. In this worst-case sequential setting, the inference time for \mech is 0.925 ms. Importantly, each inference corresponds to a trace window of 140 ms; thus, the inference overhead is negligible, accounting for only 0.7\% of the inference window duration.

Our current implementation of \mech operates entirely at the software layer. Prior work has proposed accelerating inference using specialized hardware~\cite{he2022breakthrough}, which could further reduce the computational cost of \mech. Moreover, implementations in more performant languages (e.g., Rust or C) could further decrease latency. Finally, our reported inference time for \mech reflects a pessimistic worst-case assumption in which ensemble models are evaluated serially. Parallel inference, combined with potential hardware acceleration, could substantially reduce end-to-end inference latency.

\subsection{Sequence-Learning Overheads} \label{subsec:deployment}

As discussed in \refsec{subsec:pipeline}, \mech operates in two distinct phases: offline learning and online deployment. Although sequence training is not on the critical path, in this section we empirically quantify the overhead of our \texttt{CVXPY} implementation for sequence learning in \mech. \reftab{tab:seq-learning-time} reports training times using the \texttt{CLARABEL}~\cite{goulart2024clarabel} and \texttt{SCS}~\cite{ocpb:16} solvers. The first column indicates the baselines used to construct sequences for \mech, which directly determines the size of the sequence space (see \refsec{subsec:pipeline} and \refsec{subsec:mech-perf}).

For smaller sequence spaces, \texttt{CLARABEL} converges to an optimal solution in roughly half the time required by \texttt{SCS}. However, as the sequence space grows, \texttt{SCS} becomes more efficient. \texttt{CLARABEL} is an interior-point solver that achieves higher solution accuracy by solving a KKT linear system at each iteration, a process whose cost can grow superlinearly with problem size. In contrast, \texttt{SCS} employs the alternating direction method of multipliers (ADMM), which scales more favorably for larger problems. In practice, both solvers converge to near-identical solutions, assigning comparable probabilities to the same sequences; any minor differences are likely attributable to \texttt{CLARABEL}’s higher numerical precision.

\begin{table}[t!]
    \centering
    \small
    \begin{tabular}{lclr}
        \toprule
        \mech Baselines & \# Sequences & Solver & Time (s) \\
        \midrule
        \multirow{2}{*}{\ens models} & \multirow{2}{*}{1331} & \texttt{CLARABEL} & 27.77 \\
                &  & \texttt{SCS} & 56.5 \\
        \midrule
        \multirow{2}{*}{Individual classifiers} & \multirow{2}{*}{3375} & \texttt{CLARABEL} & 1473.71 \\
                &  & \texttt{SCS} & 188 \\
        \bottomrule
    \end{tabular}
    \caption{\textbf{\mech's offline training time}}
    \label{tab:seq-learning-time}
\end{table}
\section{Conclusion and Future Directions} \label{sec:conclusion}

The persistent evolution of malware necessitates detection mechanisms that are both accurate and robust against an adversary's attempts to evade profiling. Hardware Performance Counters offer a unique vantage point for observing program behavior, but their limited concurrency forces a critical design choice: which small subset of microarchitectural events to monitor. As we have argued, relying on a single, statically optimized feature set can create systematic blind spots, limiting the detector's ability to generalize across diverse malicious and benign workloads.

In this work, we investigated the hypothesis that detection performance can be enhanced by deliberately varying the monitored feature sets over the course of a program's execution. We introduced \mech, an end-to-end mechanism that addresses the HPC concurrency constraint by segmenting execution into slices and learning an effective schedule that applies different feature sets and their corresponding classifiers across these slices. This approach explicitly trades the depth of information available per slice (temporal granularity) for breadth of behavioral coverage, countering the inherent limitations of a fixed monitoring perspective.

Our experimental findings strongly support this thesis. \mech consistently outperformed strong single-feature-set baselines, achieving substantial gains in overall detection performance while dramatically reducing the false positive rate. This demonstrates that the learned schedule successfully identifies and leverages complementary feature sets, allowing the collective decision-making process to overcome the information loss at the individual slice level. By strategically scheduling multiple feature sets over time, \mech provides a principled and effective framework to transform a fundamental architectural limitation into an opportunity for building more resilient malware detection systems.

This work opens several promising avenues for future research. First, the offline learning paradigm of \mech could be extended to support online adaptation, enabling the feature-set schedule to be dynamically adjusted based on real-time confidence estimates or detected concept drift. Second, \mech's resilience to adaptive adversarial attacks warrants further investigation, to assess whether the learned sequences inherently increase the cost and complexity of evasion or introduce new, systematic vulnerabilities. Finally, tighter hardware–software co-design could explore minimal architectural support for reducing the overhead of switching monitored events, thereby making feature-set scheduling even more efficient and practical for real-world deployment.

\bibliographystyle{plainurl}
\bibliography{references}

@inproceedings{malone2011hardware,
    title = {Are hardware performance counters a cost effective way for integrity checking of programs},
    author = {Malone, Corey and Zahran, Mohamed and Karri, Ramesh},
    booktitle = {Proceedings of the 6th ACM Workshop on Scalable Trusted Computing (STC)},
    pages = {71--76},
    year = {2011}
}

@inproceedings{zhou2018hardware,
    title = {Hardware performance counters can detect malware: {M}yth or fact?},
    author = {Zhou, Boyou and Gupta, Anmol and Jahanshahi, Rasoul and Egele, Manuel and Joshi, Ajay},
    booktitle = {Proceedings of the ACM Asia Conference on Computer and Communications Security (ASIA-CCS)},
    pages = {457--468},
    year = {2018}
}

@inproceedings{uhsadel2008exploiting,
    title = {Exploiting Hardware Performance Counters},
    author = {Uhsadel, Leif and Georges, Andy and Verbauwhede, Ingrid},
    booktitle = {Proceedings of the 5th Workshop on Fault Diagnosis and Tolerance in Cryptography (FDTC)},
    pages = {59--67},
    year = {2008}
}

@inproceedings{mucci1999papi,
    title = {{PAPI: A} portable interface to hardware performance counters},
    author = {Mucci, Philip J and Browne, Shirley and Deane, Christine and Ho, George},
    booktitle = {Proceedings of the Department of Defense HPCMP Users Group Conference (HPCMP)},
    volume = {710},
    year = {1999}
}

@inproceedings{weaver2008can,
    title = {Can hardware performance counters be trusted?},
    author = {Weaver, Vincent M and McKee, Sally A},
    booktitle = {Proceedings of the IEEE International Symposium on Workload Characterization (IISWC)},
    pages = {141--150},
    year = {2008}
}

@article{basu2019theoretical,
    title = {A theoretical study of hardware performance counters-based malware detection},
    author = {Basu, Kanad and Krishnamurthy, Prashanth and Khorrami, Farshad and Karri, Ramesh},
    journal = {IEEE Transactions on Information Forensics and Security},
    volume = {15},
    pages = {512--525},
    year = {2020},
}

@inproceedings{bahador2014hpcmalhunter,
    title = {HPCMalHunter: Behavioral malware detection using hardware performance counters and singular value decomposition},
    author = {Bahador, Mohammad Bagher and Abadi, Mahdi and Tajoddin, Asghar},
    booktitle = {Proceedings of the 4th International Conference on Computer and Knowledge Engineering (ICCKE)},
    pages = {703--708},
    year = {2014}
}

@article{ganfure2022deepware,
    title = {{DeepWare: I}maging performance counters with deep learning to detect ransomware},
    author = {Ganfure, Gaddisa Olani and Wu, Chun-Feng and Chang, Yuan-Hao and Shih, Wei-Kuan},
    journal = {IEEE Transactions on Computers},
    volume = {72},
    number = {3},
    pages = {600--613},
    year = {2022}
}

@article{kadiyala2020hardware,
    title = {Hardware performance counter-based fine-grained malware detection},
    author = {Kadiyala, Sai Praveen and Jadhav, Pranav and Lam, Siew-Kei and Srikanthan, Thambipillai},
    journal = {ACM Transactions on Embedded Computing Systems (TECS)},
    volume = {19},
    number = {5},
    pages = {1--17},
    year = {2020}
}

@inproceedings{alam2019ratafia,
    title = {{RATAFIA: R}ansomware analysis using time and frequency informed autoencoders},
    author = {Alam, Manaar and Bhattacharya, Sarani and Dutta, Swastika and Sinha, Sayan and Mukhopadhyay, Debdeep and Chattopadhyay, Anupam},
    booktitle = {Proceedings of the IEEE International Symposium on Hardware Oriented Security and Trust (HOST)},
    pages = {218--227},
    year = {2019}
}

@inproceedings{sayadi2018customized,
    title = {Customized machine learning-based hardware-assisted malware detection in embedded devices},
    author = {Sayadi, Hossein and Makrani, Hosein Mohammadi and Randive, Onkar and PD, Sai Manoj and Rafatirad, Setareh and Homayoun, Houman},
    booktitle = {Proceedings of the 17th IEEE International Conference On Trust, Security And Privacy In Computing And Communications / 12th IEEE International Conference On Big Data Science And Engineering (TrustCom/BigDataSE)},
    pages = {1685--1688},
    year = {2018}
}

@inproceedings{guthaus2001mibench,
    title = {{MiBench: A} free, commercially representative embedded benchmark suite},
    author = {Guthaus, MR and Ringenberg, JS and Ernst, D and Austin, TM and Mudge, T and Brown, RB},
    booktitle = {Proceedings of the 4th Annual IEEE International Workshop on Workload Characterization (WWC)},
    pages = {3--14},
    year = {2001}
}

@inproceedings{mcvoy1996lmbench,
    title = {{LMbench: P}ortable Tools for Performance Analysis},
    author = {McVoy, Larry W. and Carl Staelin},
    booktitle = {Proceedings of the USENIX Annual Technical Conference (ATC)},
    pages = {279--294},
    year = {1996},
}

@misc{malwarebazaar,
  title        = {MalwareBazaar: Open Malware Sharing Platform},
  key          = {MalwareBazaar},
  howpublished = {\url{https://bazaar.abuse.ch}},
  note         = {Open platform operated by abuse.ch for collecting and sharing malware samples}
}

@inproceedings{demme2013feasibility,
    author = {Demme, John and Maycock, Matthew and Schmitz, Jared and Tang, Adrian and Waksman, Adam and Sethumadhavan, Simha and Stolfo, Salvatore},
    title = {On the Feasibility of Online Malware Detection with Performance Counters},
    booktitle = {Proceedings of the 40th Annual International Symposium on Computer Architecture (ISCA)},
    pages = {559--570},
    year = {2013}
}

@inproceedings{bucek2018spec,
  title={SPEC CPU2017: Next-generation compute benchmark},
  author={Bucek, James and Lange, Klaus-Dieter and v. Kistowski, J{\'o}akim},
  booktitle={Companion of the 2018 ACM/SPEC International Conference on Performance Engineering},
  pages={41--42},
  year={2018}
}

@misc{unixbench,
  title        = {UnixBench: The BYTE UNIX Benchmark Suite},
  author       = {Smith, Ian and contributors},
  howpublished = {\url{https://github.com/kdlucas/byte-unixbench}},
  note         = {Benchmark suite providing basic performance indicators for Unix-like systems}
}

@inproceedings{makrani2022accelerated,
    title = {Accelerated Machine Learning for On-Device Hardware-Assisted Cybersecurity in Edge Platforms},
    author = {Makrani, Hosein Mohammadi and He, Zhangying and Rafatirad, Setareh and Sayadi, Hossein},
    booktitle = {Proceedings of the 23rd International Symposium on Quality Electronic Design (ISQED)},
    pages = {77--83},
    year = {2022}
}

@inproceedings{sayadi20192smart,
    title = {2{S}mart: {A} two-stage machine learning-based approach for run-time specialized hardware-assisted malware detection},
    author = {Sayadi, Hossein and Makrani, Hosein Mohammadi and Dinakarrao, Sai Manoj Pudukotai and Mohsenin, Tinoosh and Sasan, Avesta and Rafatirad, Setareh and Homayoun, Houman},
    booktitle = {Proceedings of the 19th Design, Automation \& Test in Europe Conference \& Exhibition (DATE)},
    pages = {728--733},
    year = {2019}
}

@inproceedings{sayadi2018ensemble,
    title = {Ensemble learning for effective run-time hardware-based malware detection: {A} comprehensive analysis and classification},
    author = {Sayadi, Hossein and Patel, Nisarg and Sasan, Avesta and Rafatirad, Setareh and Homayoun, Houman},
    booktitle = {Proceedings of the 55th Annual Design Automation Conference (DAC)},
    pages = {1--6},
    year = {2018}
}

@inproceedings{he2021machine,
    title = {When machine learning meets hardware cybersecurity: {D}elving into accurate zero-day malware detection},
    author = {He, Zhangying and Miari, Tahereh and Makrani, Hosein Mohammadi and Aliasgari, Mehrdad and Homayoun, Houman and Sayadi, Hossein},
    booktitle = {Proceedings of the 22nd International Symposium on Quality Electronic Design (ISQED)},
    pages = {85--90},
    year = {2021}
}

@inproceedings{he2022breakthrough,
    title = {Breakthrough to adaptive and cost-aware hardware-assisted zero-day malware detection: {A} reinforcement learning-based approach},
    author = {He, Zhangying and Makrani, Hosein Mohammadi and Rafatirad, Setareh and Homayoun, Houman and Sayadi, Hossein},
    booktitle = {Proceedings of the 40th International Conference on Computer Design (ICCD)},
    pages = {231--238},
    year = {2022}
}

@misc{clamav,
  title        = {ClamAV Antivirus Toolkit},
  key          = {ClamAv},
  howpublished = {\url{https://www.clamav.net/}},
  note         = {Open-source antivirus engine by Cisco Talos}
}

@manual{dbench,
  title        = {Dbench File System Benchmark},
  author       = {Tridgell, Andrew},
  year         = {2000},
  howpublished = {\url{https://linux.die.net/man/1/dbench}},
  note         = {File system benchmark tool}
}

@misc{phoronix,
  title        = {Phoronix Test Suite},
  author       = {Larabel, Michael},
  howpublished = {\url{https://github.com/phoronix-test-suite/phoronix-test-suite/}},
  year         = {2024},
  note         = {Open-source, cross-platform benchmarking and testing framework}
}

@misc{sysbench,
  title        = {Sysbench: Scriptable Database and System Performance Benchmark},
  author       = {Kopytov, Alexey},
  howpublished = {\url{https://github.com/akopytov/sysbench}},
  note         = {Cross-platform benchmark tool for CPU, memory, and database workloads}
}

@misc{clickbench,
  title        = {ClickBench: A Benchmark for Analytical Databases},
  author       = {ClickHouse},
  howpublished = {\url{https://github.com/ClickHouse/ClickBench}},
  note         = {Public analytical database benchmark and dataset}
}

@misc{memcached,
  title        = {Memcached},
  author       = {Fitzpatrick, Brad and contributors},
  howpublished = {\url{http://memcached.org/}},
  note         = {High-performance distributed memory object caching system}
}

@misc{memtier,
  title        = {memtier\_benchmark: High-Throughput Benchmarking Tool for Redis and Memcached},
  author       = {Redis Ltd.},
  howpublished = {\url{https://github.com/RedisLabs/memtier_benchmark}},
  note         = {Command-line load generation and benchmarking tool for Redis and Memcached}
}

@misc{redis,
  title        = {Redis: In-Memory Data Structure Store},
  author       = {Sanfilippo, Salvatore and Redis Community},
  howpublished = {\url{https://redis.io/}},
  note         = {Open-source, in-memory key-value store}
}

@misc{stress-ng,
  title        = {{Stress-ng}: System Load and Stress Testing Tool},
  author       = {King, Colin Ian},
  howpublished = {\url{https://github.com/ColinIanKing/stress-ng}},
  note         = {Tool to load and stress test a computer system}
}

@INPROCEEDINGS{cortexsuite,
  author={Thomas, Shelby and Gohkale, Chetan and Tanuwidjaja, Enrico and Chong, Tony and Lau, David and Garcia, Saturnino and Bedford Taylor, Michael},
  booktitle={2014 IEEE International Symposium on Workload Characterization (IISWC)}, 
  title={CortexSuite: A synthetic brain benchmark suite}, 
  year={2014},
  volume={},
  number={},
  pages={76-79}
}

@inproceedings{sakalis2016splash,
  title={Splash-3: A properly synchronized benchmark suite for contemporary research},
  author={Sakalis, Christos and Leonardsson, Carl and Kaxiras, Stefanos and Ros, Alberto},
  booktitle={2016 IEEE International Symposium on Performance Analysis of Systems and Software (ISPASS)},
  pages={101--111},
  year={2016},
  organization={IEEE}
}

@article{zaharia2016apache,
  title={Apache spark: a unified engine for big data processing},
  author={Zaharia, Matei and Xin, Reynold S and Wendell, Patrick and Das, Tathagata and Armbrust, Michael and Dave, Ankur and Meng, Xiangrui and Rosen, Josh and Venkataraman, Shivaram and Franklin, Michael J and others},
  journal={Communications of the ACM},
  volume={59},
  number={11},
  pages={56--65},
  year={2016}
}

@inproceedings{Bienia2008PARSEC,
  author = {Bienia, Christian and Kumar, Sanjeev and Singh, Jaswinder Pal and Li, Kai},
  title = {The PARSEC benchmark suite: characterization and architectural implications},
  booktitle = {Proceedings of the 17th international conference on Parallel architectures and compilation techniques (PACT)},
  year = {2008},
  pages = {72--81}
}

@misc{incus,
  title        = {Incus: System Container and Virtual Machine Manager},
  author       = {LXC Project},
  howpublished = {\url{https://linuxcontainers.org/incus/introduction/}},
  note         = {Modern, secure container and virtual machine manager}
}

@article{dhariyal2023scalable,
  title={Scalable classifier-agnostic channel selection for multivariate time series classification},
  author={Dhariyal, Bhaskar and Le Nguyen, Thach and Ifrim, Georgiana},
  journal={Data Mining and Knowledge Discovery},
  volume={37},
  number={2},
  pages={1010--1054},
  year={2023},
  publisher={Springer}
}

@inproceedings{arp2022and,
  title={Dos and don'ts of machine learning in computer security},
  author={Arp, Daniel and Quiring, Erwin and Pendlebury, Feargus and Warnecke, Alexander and Pierazzi, Fabio and Wressnegger, Christian and Cavallaro, Lorenzo and Rieck, Konrad},
  booktitle={Proceedings of the 31st USENIX Security Symposium (USENIX Security)},
  pages={3971--3988},
  year={2022}
}

@misc{Linux,
    key={Linux},
    title={Performance Counters for Linux},
    year={2010},
    howpublished={\url{https://lwn.net/Articles/310176/}}
}

@inproceedings{pendlebury2019tesseract,
  title={{TESSERACT}: Eliminating experimental bias in malware classification across space and time},
  author={Pendlebury, Feargus and Pierazzi, Fabio and Jordaney, Roberto and Kinder, Johannes and Cavallaro, Lorenzo},
  booktitle={Proceedings of the 28th USENIX Security Symposium (USENIX Security)},
  pages={729--746},
  year={2019}
}

@article{khasawneh2018ensemblehmd,
  title={EnsembleHMD: Accurate hardware malware detectors with specialized ensemble classifiers},
  author={Khasawneh, Khaled N and Ozsoy, Meltem and Donovick, Caleb and Abu-Ghazaleh, Nael and Ponomarev, Dmitry},
  journal={IEEE Transactions on Dependable and Secure Computing},
  volume={17},
  number={3},
  pages={620--633},
  year={2018},
  publisher={IEEE}
}

@article{diamond2016cvxpy,
  author  = {Steven Diamond and Stephen Boyd},
  title   = {{CVXPY}: {A} {P}ython-embedded modeling language for convex optimization},
  journal = {Journal of Machine Learning Research},
  year    = {2016},
  volume  = {17},
  number  = {83},
  pages   = {1--5},
}

@misc{netflix-perf1,
  title={Seeing Through Hardware Counters: A Journey to Threefold Performance Increase},
  author={Filanovsky, Vadim and Sane, Harshad},
  year={2022},
  howpublished={\url{https://netflixtechblog.com/seeing-through-hardware-counters-a-journey-to-threefold-performance-increase-2721924a2822}},
}

@misc{netflix-perf2,
  title={{Netflix} Optimizes {Amazon} Instance Performance, and Reduces Costs, Using {Intel} {Xeon} Processors and {Intel} Analysis Tools},
  author={{Intel Corporation} and {Netflix, Inc.}},
  year={2023},
  howpublished={\url{https://www.intel.com/content/dam/www/central-libraries/us/en/documents/2023-11/netflix-performance-case-study-112023.pdf}},
}

@inproceedings{google-perf-vm1,
  author={Kanev, Svilen and Darlington, Juan P. and Hazelwood, Kim and Hendrix, Paul and Hintze, Eduard and Li, Parthasarathy Ranganathan and et al.},
  title={Profiling a Warehouse-Scale Computer},
  booktitle={Proceedings of the 42nd Annual International Symposium on Computer Architecture (ISCA)},
  year= {2015},
  pages={158--169}
}

@misc{google-perf-vm2,
  title={HardwarePerfCounter: Libraries and Utilities for Sampling Hardware Performance Counters},
  author={{Google LLC}},
  year={2022},
  howpublished={\url{https://github.com/google/hardware-perfcounter}},
}

@article{anderson2018ember,
  title         = {EMBER: An Open Dataset for Training Static PE Malware Machine Learning Models},
  author        = {Anderson, Hyrum S. and Roth, Phil},
  journal       = {arXiv preprint arXiv:1804.04637},
  year          = {2018}
}

@article{or2019dynamic,
  title={Dynamic malware analysis in the modern era—A state of the art survey},
  author={Or-Meir, Ori and Nissim, Nir and Elovici, Yuval and Rokach, Lior},
  journal={ACM Computing Surveys (CSUR)},
  volume={52},
  number={5},
  pages={1--48},
  year={2019}
}

@article{goulart2024clarabel,
  title={Clarabel: An interior-point solver for conic programs with quadratic objectives},
  author={Goulart, Paul J and Chen, Yuwen},
  journal={arXiv preprint arXiv:2405.12762},
  year={2024}
}

@article{ocpb:16,
    author       = {Brendan O'Donoghue and Eric Chu and Neal Parikh and Stephen Boyd},
    title        = {Conic Optimization via Operator Splitting and Homogeneous Self-Dual Embedding},
    journal      = {Journal of Optimization Theory and Applications},
    month        = {June},
    year         = {2016},
    volume       = {169},
    number       = {3},
    pages        = {1042-1068}
}

@inproceedings{cheng2023feasibility,
  title={On the feasibility of malware unpacking via hardware-assisted loop profiling},
  author={Cheng, Binlin and Leal, Erika A and Zhang, Haotian and Ming, Jiang},
  booktitle={Proceedings of the 32nd USENIX Security Symposium (USENIX Security)},
  pages={7481--7498},
  year={2023}
}

\end{document}